\documentclass[%
 reprint,
 amsmath,amssymb,
 aps,
]{revtex4-2}

\usepackage{color}
\usepackage{bbm}
\usepackage{mathrsfs}
\usepackage{graphicx}
\usepackage{dcolumn}
\usepackage{bm}

\begin{document}

\title{General Grad-Shafranov Equation}

\author{Ye Shen}
\email{shenye199594@fzu.edu.cn}
\affiliation{College of Physics and Information Engineering,\\ Fuzhou University, Fuzhou 350108, Fujian, P. R. China}

\date{\today}

\begin{abstract}

To effectively describe the plasma with strong magnetic field, the force-free electrodynamics was introduced, within which the Grad-Shafranov equation plays the key role. The Grad-Shafranov equation governs the global structure of a electromagnetic field in equilibrium with symmetries. It is widely applicable in an amount of scenarios, such as the tokamak, the solar corona, the magnetosphere of Earth, neutron star and black hole, etc. However, in different situations, the Grad-Shafranov equation is expressed differently, and the derivations might be complicated. In this work, via the language of differential form, we provide a general expression of Grad-Shafranov equation, from which the expression in any specific situation can be quickly obtained. Additionally, we present a Lagrangian density for a scalar field whose on-shell condition is precisely the Grad-Shafranov equation.

\end{abstract}

\maketitle


\section{Introduction}
\label{sec:intro}

In plasma physics and astrophysics, strong electromagnetic fields are frequently encountered. Generally, the energy of a plasma can be separated into fluid and electromagnetic components. Whenever the electromagnetic part dominates, the force-free electrodynamics framework, which imposes $F\cdot j=0$ (where $F$ is the electromagnetic 2-form and $j$ is the current 4-vector), becomes effective \cite{Mizuno:2025mog}. 

To analyze the global structure of stationary force-free magnetic fields with symmetries, Grad and Shafranov introduced the stream equation, now referred to be the Grad-Shafranov (G–S) equation, by directly examining the force-free condition $F\cdot j=0$ \cite{Grad1958,Shafranov1966}. It is a second order partial differential equation (PDE) of a scalar potential, whose first order derivatives turn out to be the components of magnetic strength. This breakthrough helps people to theoretically handle the structure of Earth's magnetosphere \cite{ANTONOVA1997,Uzdensky:2004quu,Sonnerup2006,Chen2018} and the solar coronal mass ejection process \cite{Forbes1995,Lin2000,Reeves2005,Lin2006}. It also boosts the development of theories describing plasma in tokamak \cite{Atanasiu2004,abel2013,Li2021,Pentland2025}.

The discoveries of pulsar \cite{Hewish1,Hewish2,Gold1968,Pacini1968,Goldreich1969,Brennan:2013ppa} and quasar \cite{Hewish3,Schmidt1963}, both thought to be powered by strong magnetic pressure near neutron stars or black holes, encouraged the extension of force-free electrodynamics to curved spacetimes. Via the covariant language in general relativity, stationary and axisymmetric fields near a rotating compact object were studied \cite{BZ,Uzdensky:2003cy,Uzdensky:2004qu,Komissarov:2004ms,Gralla:2015vta,Gelles:2024tpz,Song:2025mhj}, including the preliminary form of the G-S equation in Kerr spacetime \cite{Nitta:1991qx,Ioka:2003dd,Contopoulos:2012py,Nathanail:2014aua}. To determine the global structure of magnetic fields near a compact object, significant efforts have been devoted to solving the Grad–Shafranov equation in Kerr spacetime through perturbative method \cite{Brennan:2013kea,Armas:2020mio,Camilloni:2022kmx} and numerical simulations \cite{McKinney2006-1,McKinney2006-2,Mahlmann2018,Kim:2024mau}. One of the most famous results, the split monopole \cite{Michel1973,Bogovalov:1999vg,Kalapotharakos2012,Contopoulos:2013ota}, was proven by magnetohydrodynamics and particle-in-cells simulations to be effective to estimate the magnetosphere of a Kerr black hole \cite{Komissarov:2003tg,McKinney:2004ka,SashaNote,Ruiz:2023hit,Xi:2024sxb,Yang:2024kpz,Vos:2024loa}. All these contributions help us uncover the mysteries of strong magnetic field in complex astrophysical environments.

The preliminary form of G-S equation in Kerr spacetime is reliable but algebraically cumbersome. Through in-depth analyses of degenerate fields \cite{Uchida1,Uchida2,Uchida3,Uchida4,Uchida5,Komissarov:2002my} and the language of differential form \cite{Gralla:2014yja}, force-free electrodynamics in Kerr spacetime, along with the G-S equation, could be expressed in a more concise way \cite{Gralla:2014yja,Camilloni:2022kmx}. It is predictable that researchers are prepared to extend the description of force-free electrodynamics to more complex situations, for which the specific expressions of the G-S equation will be necessary. 

In this work, a general and compact expression of G-S equation is derived, as shown in Eq.~\eqref{eq:G-S-2}, which is capable of describing the force-free electromagnetic field in an arbitrary 4D manifold with symmetries along any pair of commuting 4-vector fields. Starting from Eq.~\eqref{eq:G-S-2}, all well-known forms of the G-S equation can be quickly recovered by inputting the information of relevant metric and symmetries, thereby confirming the correctness of this general formula. Moreover, the G-S equation in any specific case that was not previously discussed can also be readily obtained. Furthermore, we argue that the G-S equation could be interpreted as the on-shell condition for a scalar field whose dynamics is contained within the Lagrangian density presented in Eq.~\eqref{eq:lagrangian}. From this Lagrangian density, the property of light surface, typically treated as a natural boundary separating different parts of the force-free field, could be directly seen.

This paper is organized as follows. In Sect.~\ref{sec:G-S}, the derivation of Eq.~\eqref{eq:G-S-2}, the general G-S equation, is presented. In Sect.~\ref{sec:example}, to verify the correctness of Eq.~\eqref{eq:G-S-2}, we demonstrate how this general equation reduces to several well-known forms of the G-S equation. In Sect.~\ref{sec:dynamics}, we provide the Lagrangian density and briefly discuss the dynamics it encodes. We make a summary in Sect.~\ref{sec:sum} and discuss some minor details. To assist readers in following our calculations, we provide basic information about the differential forms in Appen.~\ref{sec:def}. The properties of degenerate field and comoving 1-form are discussed in Appen.~\ref{sec:degenerate} and \ref{sec:comoving}, respectively. In Appen.~\ref{sec:unique}, we discuss the sufficient conditions under which the general G-S equation admits a unique solution given appropriate boundary conditions. Some equations used in the main text are proved in Appen.~\ref{sec:tags}.

\section{Grad-Shafranov Equation}
\label{sec:G-S}

\subsection{Basic Settings}
\label{sec:basic}

In general, the G-S equation describes a force-free electromagnetic field in a 4D manifold which contains two symmetries with respect to two linearly independent, commuting 4-vector fields (typically one timelike and one spacelike). Denote this pair of vectors as $e_0$ and $e_3$. Because they commute, there exists a coordinate basis $\{\tau,\chi,\xi,\zeta\}$ such that $e_0^{\mu}\equiv\partial_{\tau}^{\mu}$, $e_3^{\mu}\equiv\partial_{\chi}^{\mu}$ \cite{Misner:1973prb}. The electromagnetic field then satisfies:
\begin{equation}
    \mathcal{L}_{e_0}F=\mathcal{L}_{e_3}F=0
    \label{eq:symmetries}
\end{equation}
where $\mathcal{L}_X$ denotes the Lie derivative with respect to the vector $X$. 

Because the force-free field is degenerate, it could be decomposed using a pair of scalars $\{\phi_1,\phi_2\}$, called Euler potentials, such that $F={\rm d}\phi_1\wedge{\rm d}\phi_2$ \cite{Uchida1,Uchida2}. Moreover, the force-free condition can be written in terms of the Euler potentials as (read Appen.~\ref{sec:degenerate} for details):
\begin{equation}
    {\rm d}\phi_1\wedge{\rm d}\ast F=0~~~{\rm and}~~~{\rm d}\phi_2\wedge{\rm d}\ast F=0
    \label{eq:ff}
\end{equation}

The complete integrability of ${\rm d}\phi_1$ and ${\rm d}\phi_2$ in a 4D manifold indicates the existence of a pair of linearly independent smooth vector fields that annihilate both ${\rm d}\phi_1$ and ${\rm d}\phi_2$ (i.e., $X$ annihilates ${\rm d}\phi_1$ means $X\cdot{\rm d}\phi_1= 0$ at every points of the manifold) \cite{Chernbook}. Traditionally, attentions are focused on cases where either $e_0$ or $e_3$, such as $e_0=\partial_t$, serves as one of the vector fields annihilating ${\rm d}\phi_1$ and ${\rm d}\phi_2$. While a more general situation is that one could find a linear combination $L\left(e_0,e_3\right):=Ae_0+Be_3$ satisfying:
\begin{equation}
    {\rm d}\phi_1\cdot L\left(e_0,e_3\right)={\rm d}\phi_2\cdot L\left(e_0,e_3\right) = 0
    \label{eq:fro}
\end{equation}
where $\{A,B\}$ are a pair of scalars.

Conclusively, the G-S equation is the equation satisfied by a closed, degenerate field in a 4D manifold obeying Eq.~\eqref{eq:symmetries}, Eq.~\eqref{eq:ff} and Eq.~\eqref{eq:fro}.

\subsection{General Equation}
\label{sec:equation}

With the basic settings discussed above, let us derive the general G-S equation, which is applicable for any kind of 4D manifold with any pair of commuting symmetries. starting from Eq.~\eqref{eq:symmetries}, the Cartan's magic formula gives (illustrating with $e_3$ only):
\begin{equation}
    0=\mathcal{L}_{e_3}F=e_3\cdot {\rm d}F+{\rm d}\left(e_3\cdot F\right)
                      ={\rm d}\left(e_3\cdot F\right)
    \label{eq:Cartan}
\end{equation}
The equation above indicates that both $e_3\cdot F$ and $e_0\cdot F$ are closed 1-forms. Subsequently, one can find a pair of scalars $\{f,g\}$ satisfying $e_3\cdot F={\rm d}f$ and $e_0\cdot F={\rm d}g$. Intuitively, $\{f,g\}$ should be closely related to the Euler potentials. To figure out this relation, we compute:
\begin{equation}
    {\rm d}f \wedge {\rm d}g = \left(e_3\cdot F\right)\wedge\left(e_0\cdot F\right)
                 = \left(e_0\cdot e_3\cdot F\right)F
    \label{eq:dfdg}
\end{equation}
The second step is proved in Appen.~\ref{sec:prove-eq4}. The factor $\left(e_0\cdot e_3\cdot F\right)$ is a scalar. Furthermore, Cartan's magic formula implies:
\begin{equation}
    \begin{aligned}
        {\rm d}\left(e_0\cdot e_3\cdot F\right)
        &=\mathcal{L}_{e_0}\left(e_3\cdot F\right)-e_0\cdot {\rm d}\left(e_3\cdot F\right) \\
        &=e_3\cdot \left(\mathcal{L}_{e_0}F\right) + \left[e_0,e_3\right]\cdot F \\
        &=0
    \end{aligned}
\end{equation}
Hence the factor $\left(e_3\cdot e_0\cdot F\right)$ is actually a constant. It can be shown that one cannot find a linear combination of $e_0$ and $e_3$ to annihilate both ${\rm d}\phi_1$ and ${\rm d}\phi_2$ when $\left(e_3\cdot e_0\cdot F\right)$ is nonzero \cite{Gralla:2014yja}. For the satisfaction of Eq.~\eqref{eq:fro}, we therefore consider only the case of ${\rm d}f\wedge{\rm d}g=0$. This implies:
\begin{equation}
    {\rm d}g\equiv e_0\cdot F=\left(e_0\cdot {\rm d}\phi_1\right){\rm d}\phi_2
    -\left(e_0\cdot {\rm d}\phi_2\right){\rm d}\phi_1 
    = g'(f){\rm d}f
    \label{eq:dg}
\end{equation}
We may then conveniently choose $\phi_1=f$ such that:
\begin{equation}
    \begin{aligned}
        & e_0\cdot{\rm d}\phi_1=0~~~,~~~e_0\cdot{\rm d}\phi_2=-g'\left(\phi_1\right) \\
        & e_3\cdot{\rm d}\phi_1=0~~~,~~~e_3\cdot{\rm d}\phi_2=1 
    \end{aligned}
    \label{eq:case1}
\end{equation}
Eq.~\eqref{eq:fro} is then satisfied with $L\left(e_0,e_3\right)=e_0+g'(\phi_1)e_3$. For a nonzero ${\rm d}f$ (the case ${\rm d}f=0$ is discussed in Sect.~\ref{sec:sum}), the relations in Eq.~\eqref{eq:case1} restrict the form of the Euler potentials to:
\begin{equation}
    \phi_1=\psi~~~,~~~\phi_2=\eta+\chi-\omega\tau
    \label{EP-case1}
\end{equation}
for $\psi$ and $\eta$ being functions of $\{\xi,\zeta\}$ only. With Eq.~\eqref{EP-case1}, the electrodynamic 2-form could be expressed as:
\begin{equation}
    F={\rm d}\phi_1\wedge{\rm d}\phi_2
    ={\rm d}\psi\wedge{\rm d}\eta + {\rm d}\psi\wedge\mathcal{Z}
    \label{eq:F-case1}
\end{equation}
where $\mathcal{Z}={\rm d}\chi-\omega{\rm d}\tau$. Here, we denote $\omega\left(\psi\right) =g'\left(\psi\right)$. The physical interpretations of $\mathcal{Z}$ and $\omega$ is discussed in Appen.~\ref{sec:comoving}.

Now let us examine the force-free condition shown in Eq.~\eqref{eq:ff}. Since $\psi$ and $\eta$ depend only on $\{\xi,\zeta\}$, the Hodge dual of ${\rm d}\psi\wedge{\rm d}\eta$ must be proportional to ${\rm d}\tau\wedge{\rm d}\chi$. One might as well set the ratio to be $G$ such that:
\begin{equation}
    \ast\left({\rm d}\psi\wedge{\rm d}\eta\right)\equiv G{\rm d}\tau\wedge{\rm d}\chi
    \label{eq:G}
\end{equation}
The first equation in Eq.~\eqref{eq:ff} then becomes:
\begin{equation}
    \begin{aligned}
    0={\rm d}\psi\wedge{\rm d}\ast F=
    &{\rm d}\psi\wedge{\rm d}G\wedge{\rm d}\tau\wedge{\rm d}\chi \\
    +&{\rm d}\psi\wedge{\rm d}\ast\left({\rm d}\psi\wedge{\rm d}\chi\right) \\
    -&{\rm d}\psi\wedge{\rm d}\ast\left(\omega{\rm d}\psi\wedge{\rm d}\tau\right)
    \end{aligned}
    \label{eq:ff1-case1}
\end{equation}
The second and third terms on the right-hand-side vanish identically(see Appen.~\ref{sec:tags}). As a consequence, the above equation reduces to ${\rm d}\psi\wedge{\rm d}G$=0, which implies $G=G(\psi)$. 

Now the current 3-form could be written as:
\begin{equation}
    J={\rm d}\ast F=G'{\rm d}\psi\wedge{\rm d}\tau\wedge{\rm d}\chi
    +{\rm d}\ast\left({\rm d}\psi\wedge\mathcal{Z}\right)
    \label{eq:J}
\end{equation}
where $G'\equiv dG/d\psi$. Substitute Eq.~\eqref{eq:J} into the second equation of Eq.~\eqref{eq:ff}. Most terms given by ${\rm d}\phi_2\wedge{\rm d}\ast F$ actually vanish (see Appen.~\ref{sec:tags}). The second equation of Eq.~\eqref{eq:ff} then becomes:
\begin{equation}
    G'{\rm d}\eta\wedge{\rm d}\psi\wedge{\rm d}\tau\wedge{\rm d}\chi
    + \mathcal{Z}\wedge{\rm d}\ast\left({\rm d}\psi\wedge\mathcal{Z}\right) =0
    \label{eq:G-S-1}
\end{equation}
Using Eq.~\eqref{eq:G}, we have:
\begin{equation}
    \begin{aligned}
        {\rm d}\eta\wedge{\rm d}\psi 
        &= (-1)^{s+4}\ast\ast\left({\rm d}\eta\wedge{\rm d}\psi\right) \\
        &= (-1)^{s+1}\ast\left(G{\rm d}\tau\wedge{\rm d}\chi\right)
    \end{aligned}
\end{equation}
where $s$ is the negative index of inertia of the metric tensor $[g_{\mu\nu}]$ (see Eq.~\eqref{eq:hodge-square}). The first term in Eq.~\eqref{eq:G-S-1} therefore becomes:
\begin{equation}
    (-1)^{s+1}G'\ast\left(G{\rm d}\tau\wedge{\rm d}\chi\right)\wedge{\rm d}\tau\wedge{\rm d}\chi
    =\frac{(-1)^{s+1}}{4}\frac{dG^2}{d\psi}\left|{\rm d}\tau\wedge{\rm d}\chi\right|^2 \epsilon
    \label{eq:1term}
\end{equation}
with $\epsilon$ the metric-compatible volume element defined in Eq.~\eqref{eq:volume1} or Eq.~\eqref{eq:volume2}. We have applied Eq.~\eqref{eq:hodge_prop} in Eq.~\eqref{eq:1term}. The second term in Eq.~\eqref{eq:G-S-1}, being a 4-form in a 4D manifold, is also proportional to the metric-compatible volume element as well. It is not hard to figure out that:
\begin{equation}
    \mathcal{Z}\wedge{\rm d}\ast\left({\rm d}\psi\wedge\mathcal{Z}\right)
    =-2\mathcal{Z}_{\alpha}\nabla_{\beta}\left(\mathcal{Z}^{[\alpha}\partial^{\beta ]}\psi\right)\epsilon
    \label{eq:2term}
\end{equation}
Combining Eq.~\eqref{eq:G-S-1}, Eq.~\eqref{eq:1term} and Eq.~\eqref{eq:2term}, the second equation in Eq.~\eqref{eq:ff} finally turns out to be:
\begin{equation}
    \begin{aligned}
        &\frac{(-1)^s}{8}\frac{dG^2}{d\psi}\left|{\rm d}\tau\wedge{\rm d}\chi\right|^2+\mathcal{Z}_{\alpha}\nabla_{\beta}\left(\mathcal{Z}^{[\alpha}\partial^{\beta]}\psi\right)=0 \\
        &~~~~~~~~~~~~~~~~\left({\rm with}~~\mathcal{Z}={\rm d}\chi-\omega{\rm d}\tau\right)
    \end{aligned}
    \label{eq:G-S-2}
\end{equation}
The equation above is just the general G-S equation we seek. It governs the relation among the functions $\{\psi,G,\omega\}$, which charaterize the global structure of a force-free electromagnetic field in a 4D manifold containing two symmetries with respect to $\partial_{\tau}$ and $\partial_{\chi}$. Because both $G$ and $\omega$ are functions of $\psi$, Eq.~\eqref{eq:G-S-2} is in fact a second order PDE of $\psi$.

\subsection{Further Simplifications}
\label{sec:simplification}

Generally speaking, one is concerned with the global structure of a field in a 4D Lorenzian manifold, for which $s=1$ in Eq.~\eqref{eq:G-S-2}. Magnetically dominated configurations, which are thought to be much more common in plasma and astrophysics, satisfy $F^2=2\left(B^2-E^2\right)=\big|{\rm d}\phi_1\big|^2\big|{\rm d}\phi_2\big|^2>0$ \cite{Uchida1,Uchida2}, hence both ${\rm d}\phi_1$ and ${\rm d}\phi_2$ must be spacelike 1-form. We already know that $\phi_1\equiv\psi$ depends on ${\xi,\zeta}$ only, which indicates the 1-form ${\rm d}\psi$ is a linear combination of ${\rm d}\xi$ and ${\rm d}\zeta$. To guarantee the spacelike property of ${\rm d}\psi$, both ${\rm d}\xi$ and ${\rm d}\zeta$ are necessarily spacelike, which in turns implies that either ${\rm d}\tau$ or ${\rm d}\chi$ is timelike (or both of them are null). We may conveniently choose a timelike ${\rm d}\tau$ and a spacelike ${\rm d}\chi$. Then $\mathcal{Z}$ could be interpreted as the comoving 1-form and $\omega$ as the velocity of the field (see Appen.~\ref{sec:comoving} for the details).

In most cases of interest, the metric tensor, expressed in the coordinates $\{\tau,\chi,\xi,\zeta\}$, is block-diagonal: $[g_{\mu\nu}]=[g^L_{\mu_1\nu_1}]\oplus[g^R_{\mu_2\nu_2}]$. Consequently, the 4D manifold can be decomposed into a 2D submanifold (mostly Lorenzian) spanned by $\{{\rm d}\tau,{\rm d}\chi\}$ and a 2D submanifold (mostly Riemannian) spanned by $\{{\rm d}\xi,{\rm d}\zeta\}$ (see Appen.~\ref{sec:decompose}). In this case, we have $\big|{\rm d}\tau\wedge{\rm d}\chi\big|^2=-2\big|g^L\big|^{-1}$ and
\begin{equation}
    F=\pm G\sqrt{\Bigg|\frac{g^R}{g^L}\Bigg|}{\rm d}\xi\wedge{\rm d}\zeta+{\rm d}\psi\wedge\mathcal{Z}
    \label{eq:F-sim}
\end{equation} 
where $g^L$ and $g^R$ denote the determinants of $[g^L_{\mu_1\nu_1}]$ and $[g^R_{\mu_2\nu_2}]$, respectively. The choice of sign in the first term depends on the properties of coordinates (though it is ultimately unimportant for the G-S equation). Specifically, it should be negative for a Lorenzian submanifold spanned by $\{{\rm d}\tau,{\rm d}\chi\}$ and a Riemannian submanifold spanned by $\{{\rm d}\xi,{\rm d}\zeta\}$. From Eq.~\eqref{eq:F-sim}, one finds that $G$ is proportional to $F_{\xi\zeta}$, while the gradient of $\psi$ accounts for the remaining nonzero components of $F_{\mu\nu}$. That is why $\psi$ is often referred to as the scalar potential of the field.

Additionally, the second term on the left-hand-side of Eq.~\eqref{eq:G-S-2} satisfies:
\begin{equation}
    \begin{aligned}
        2\mathcal{Z}_{\alpha}\nabla_{\beta}\left(\mathcal{Z}^{[\alpha}\partial^{\beta]}\psi\right)
        &=\frac{1}{\sqrt{\big|g\big|}}\mathcal{Z}_{\alpha}\partial_{\beta}\left(\sqrt{\big|g\big|}\mathcal{Z}^{\alpha}\partial^{\beta}\psi\right) \\
        &-\frac{1}{\sqrt{\big|g\big|}}\mathcal{Z}_{\alpha}\partial_{\beta}\left(\sqrt{\big|g\big|}\mathcal{Z}^{\beta}\partial^{\alpha}\psi\right)
    \end{aligned}
    \label{eq:divergence}
\end{equation}
where the second term on the right-hand-side above is identically zero when the metric tensor is block-diagonal. Therefore, focusing on a 4D Lorenzian manifold that can be decomposed into two 2D submanifolds in the coordinates $\{\tau,\chi,\xi,\zeta\}$, the general G-S equation shown in Eq.~\eqref{eq:G-S-2} reduces to:
\begin{equation}
    \begin{aligned}
        &~~~~\frac{1}{2}\frac{dG^2}{d\psi}+\sqrt{\Bigg|\frac{g^L}{g^R}\Bigg|}\mathcal{Z}_{\alpha}\partial_{\beta}\left(\sqrt{\big|g\big|}\mathcal{Z}^{\alpha}g^{\beta\gamma}\partial_{\gamma}\psi\right)=0 \\
        &\left({\rm with}~~\mathcal{Z}={\rm d}\chi-\omega{\rm d}\tau~~{\rm and}~~
        G=\pm\left|g^L/g^R\right|^{1/2}F_{\xi\zeta}\right)
    \end{aligned}
    \label{eq:G-S-4}
\end{equation}
This equation is equivalent to the Eq.~(86) in Ref.~\cite{Gralla:2014yja} and covers most situations typically encountered.

\section{From General to Special Cases}
\label{sec:example}

To verify the correctness of Eq.~\eqref{eq:G-S-2}, we demonstrate show how the well-known forms of the G-S equation in different situations can be derived from this general expression. Three cases are under consideration. First, a stationary field in  Minkowski spacetime which is symmetric along $z$-axis in the rest frame where the electric field vanishes. Second, a stationary field in Minkowski spacetime which is cylindrically symmetric in the rest frame where the electric field vanishes. Third, a stationary, axisymmetric field in Kerr spacetime in Boyer-Lindquist coordinates. All the three cases satisfy the following conditions: field is magnetically dominated, the 4D manifold is Lorenzian and can be decomposed into a 2D Lorentzian and a 2D Riemannian submanifolds. Therefore, we may start from Eq.~\eqref{eq:G-S-4} rather than Eq.~\eqref{eq:G-S-2}.

For the first case, we have: $\{\tau,\chi,\xi,\zeta\}\mapsto\{t,z,x,y\}$. The metric tensor is $[g_{\mu\nu}]={\rm diag}(-1,\mathbbm{1}_{3\times 3})$. The field is invariant along $z$-axis ($e_3=\partial_z$) and under time translations ($e_0=\partial_t$). The determinants of the metric tensors are: $g=-1$, $g^L=-1$ and $g^R=1$. From the second term on the right-hand-side of Eq.~\eqref{eq:F-sim}, we find $B_x=\partial_y\psi$, $B_y=-\partial_x\psi$. Thus $\psi$ is classically called the magnetic scalar potential. Meanwhile, from the first term on the right-hand-side of Eq.~\eqref{eq:F-sim}, we obtain $G=B_z$. We are interested in the rest frame where the electric field vanishes, corresponding to $L\left(e_0,e_3\right)=e_0$ and hence $\mathcal{Z}={\rm d}z$. Summarizing the discussions above, Eq.~\eqref{eq:G-S-4} yields the G-S equation in Cartesian coordinates:
\begin{equation}
    \frac{1}{2}\frac{dB_z^2}{d\psi}+\left(\partial_x^2+\partial_y^2\right)\psi=0
    \label{eq:G-S-carte}
\end{equation}
which is one of the most famous forms, originally derived in Ref.~\cite{Grad1958,Shafranov1966} and is applicable to studies of the solar coronal mass ejection process \cite{Forbes1995,Lin2000,Reeves2005,Lin2006}.

For the second case, the coordinates are mapped as: $\{\tau,\chi,\xi,\zeta\}\mapsto\{t,\varphi,r,z\}$. The metric tensor is $[g_{\mu\nu}]={\rm diag}(-1,r^2,1,1)$, so that $g=-r^2$, $g^L=-r^2$ and $g^R=1$. The field is invariant under time translations and the rotations about $z$-axis. Eq.~\eqref{eq:F-sim} implies $G=rB_{\varphi}$, $B_r=\partial_z\psi$ and $B_z=-\partial_r\psi$. The vanishment of electric field in the rest frame of $\partial_t$ gives $\omega=0$ and $\mathcal{Z}={\rm d}\varphi$. Taking all these into Eq.~\eqref{eq:G-S-4}, we obtain:
\begin{equation}
    \frac{1}{2}\frac{d\left(rB_{\varphi}\right)^2}{d\psi}+\partial_r^2\psi-r^{-1}\partial_r\psi+\partial_z^2\psi=0
    \label{eq:G-S-cylin}
\end{equation}
which is another well-known expression of G-S equation in flat spacetime and is frequently used to describe the magnetic field in tokamak \cite{Atanasiu2004,abel2013,Li2021,Pentland2025}.

Now we consider the Kerr spacetime in Boyer-Lindquist coordinates: $\{\tau,\chi,\xi,\zeta\}\mapsto\{t,\varphi,r,\theta\}$. The nonzero components of the metric tensor are: $g_{tt}=-(1-2Mr\Sigma^{-1})$, $g_{\varphi\varphi}=\left(r^2+a^2+2Ma^2r\Sigma^{-1}\sin^2\theta\right)\sin^2\theta$, $g_{rr}=\Sigma\Delta^{-1}$, $g_{\theta\theta}=\Sigma$ and $g_{t\varphi}=g_{\varphi t}=-2Mar\Sigma^{-1}\sin^2\theta$, where the length scales are defined by $a=J/(Mc)$, $\Sigma=r^2+a^2\cos^2\theta$ and $\Delta=r^2-2Mr+a^2$, with $M$ and $J$ denoting the mass and angular momentum of the central black hole, respectively. The determinants of the metric tensors are: $g=-\Sigma^2\sin^2\theta$, $g^L=-\Delta\sin^2\theta$ and $g^R=\Sigma^2\Delta^{-1}$. Outside the event horizon, $g^R$ is positive while both $g$ and $g^L$ are negative. The function $G=\left(\Sigma^{-1}\Delta\sin\theta\right)F_{r\theta}$, which could be denoted as $I$, internally represents the poloidal current crossing a 2D surface surrounded by a loop of fixed $r$ and $\theta$ \cite{Gralla:2014yja}. Now $\mathcal{Z}$ is called the corotating 1-form and $\omega=-\left(\partial_rA_t\right)/\left(\partial_rA_{\phi}\right)=-\left(\partial_{\theta}A_t\right)/\left(\partial_{\theta}A_{\phi}\right)$ manifests as the angular velocity of the field. In this case, Eq.~\eqref{eq:G-S-4} reduces to:
\begin{equation}
    \begin{aligned}
        \frac{\Sigma}{2\Delta\sin\theta}\frac{dI^2}{d\psi}
        &+\mathcal{Z}_{\alpha}\partial_r\left(\mathcal{Z}^{\alpha}\Delta\sin\theta\partial_r\psi\right) \\
        &+\mathcal{Z}_{\alpha}\partial_{\theta}\left(\mathcal{Z}^{\alpha}\sin\theta\partial_{\theta}\psi\right)=0
    \end{aligned}
    \label{eq:G-S-Kerr}
\end{equation}
which is equivalent to the G-S equation derived in Ref.~\cite{Camilloni:2022kmx}.

In more complicated cases, one has to start from Eq.~\eqref{eq:G-S-2}. Let us discuss the general stationary spacetime as a final example, for which the metric is not necessarily block-diagonal so that the 2+2 decomposition may not always be made. The line element takes the form of $ds^2=-N^2 dt^2 + \gamma_{ij}\left(dx^i+\beta^i dt\right)\left(dx^j+\beta^j dt\right)$, with $N$ the lapse function, $\beta^i$ the shift vector and $\gamma_{ij}$ the induced metric in 3D spatial slices. We might as well choose $\{\tau,\chi,\xi,\zeta\}\mapsto\{t,x^3,x^1,x^2\}$, with which $\left|{\rm d}\tau\wedge{\rm d}\chi\right|^2=-2\gamma^{33}/N^2$. Based on Eq.~\eqref{eq:G-S-2}, now the G-S equation should be expressed as:
\begin{equation}
    \begin{aligned}
        \frac{\gamma^{33}\sqrt{\gamma}}{4N}\frac{dG^2}{d\psi} &=
        \partial_{\kappa}\left[\mathcal{Z}^{\kappa}\left(N\sqrt{\gamma}\gamma^{3m}-\frac{\beta^3\beta^m}{N}\right)\partial_m\psi\right] \\
        -& \mathcal{Z}_{\kappa}\partial_l\left[\mathcal{Z}^{\kappa}\left(N\sqrt{\gamma}\gamma^{lm}-\frac{\beta^l\beta^m}{N}\right)\partial_m\psi\right] \\
        -& \omega\partial_{\kappa}\left(\mathcal{Z}^{\kappa}\frac{\beta^m\sqrt{\gamma}}{N}\partial_m\psi\right) 
    \end{aligned}
    \label{eq:G-S-Gen}
\end{equation}
where $\kappa=\{0,1,2,3\}$, $m=\{1,2\}$ and $l=\{1,2,3\}$. Here $\gamma^{ij}$ is defined by $\gamma^{ik}\gamma_{kj}=\delta^i_j$ and $\gamma$ denotes the determinant of $\gamma_{ij}$. We adopt in E.~\eqref{eq:G-S-Gen} that $N$, $\gamma_{ij}$ and $\beta^i$ are independent of $t$. The field, based on Eq.~\eqref{eq:F-case1}, then takes the form of:
\begin{equation}
    \begin{aligned}
        F &= G\frac{\sqrt{\gamma}}{N}\varepsilon_{ijk}\left(\gamma^{i3}dx^j\wedge dx^k - 2\beta^{[i}\gamma^{j]3}dt\wedge dx^k\right) \\
        &~~+ \partial_m\psi~dx^m \wedge \left(dx^3-\omega dt\right)
    \end{aligned}
    \label{eq:F-Gen}
\end{equation}
with $m=\{1,2\}$ and $\varepsilon_{ijk}$ the 3D Levi-Civita symbols.

\section{Dynamics}
\label{sec:dynamics}

As long as $\mathcal{Z}$ is not null (mostly satisfied), the Eq.~\eqref{eq:G-S-2} could be regarded as the on-shell condition of the scalar field $\psi$ whose Lagrangian density takes the form of:
\begin{equation}
    \mathscr{L} = \frac{1}{2}\left(\left|\rm{d}\psi\right|^2-\left|\hat{\mathcal{Z}}\cdot\rm{d}\psi\right|^2\right)
    - \frac{\left(-1\right)^{s+1}}{8}\lambda_{\tau\chi}\left|\Tilde{\mathcal{Z}}\left(G^2\right)\right|^{-1}
    \label{eq:lagrangian}
\end{equation}
where $\hat{\mathcal{Z}}\equiv \mathcal{Z}/\left|\mathcal{Z}\right|$ and $\lambda_{\tau\chi}^2\equiv -\left|\rm{d}\tau\wedge\rm{d}\chi\right|^{2}$. Here $\Tilde{\mathcal{Z}}\left(G^2\right):=\rm{d\chi}-\Tilde{\omega}\left(G^2\right)\rm{d}\tau$, with $\Tilde{\omega}\left(G^2\right)$ being equivalent to $\omega\left(\psi\right)$ but expressed as a function of $G^2$.  A more robust expression of the last term is:
\begin{equation}
    -\frac{\left(-1\right)^{s+1}}{8}\lambda_{\tau\chi}\int^{G^2}{\left|\mathcal{Z}\right|^{-2}d\Tilde{G}^2}
\end{equation}

The first term of $\mathscr{L}$ represents the kinetic energy of $\psi$, in which only the components perpendicular to $\mathcal{Z}$ contribute. While $\mathcal{Z}$ itself acts as a potential in the dynamics, as reflected in the second term. The field tends to keep in the equilibrium that minimizes $\left(-1\right)^{s+1}\left|\Tilde{\mathcal{Z}}\left(G^2\right)\right|^{-1}$. 

Let us consider the simplest case, where the manifold is Lorentzian, and the timelike $\rm{d}\tau$ and spacelike $\rm{d}\chi$ are orthogonal to each other. In this case, $\lambda_{\tau\chi}>0$. The field $\psi$ could then be viewed as massless particles confined to the hypersurfaces perpendicular to $\mathcal{Z}$, which reach a stable equilibirium provided that $\Tilde{\omega}\left(G^2\right)$ is locally maximized. Interpreting $\mathcal{Z}$ as the comoving 1-form (see Appen.~\ref{sec:comoving}), the argument above implies a higher comoving velocity corresponds to a more stable state. In other words, the motion of a force-free electromagnetic field enhances its own stability.

From Eq.~\eqref{eq:lagrangian}, it is revealed that the light surface, where $\mathcal{Z}$ becomes null, is a critical surface on which the potential diverges. Viewing $\psi$ as the massless particles moving on the hypersurfaces perpendicular to $\mathcal{Z}$, the light surface is a singularity that these particles could approach arbitrarily closely but never reach. Consequently, the light surface acts as a natural barrier that these particles cannot penetrate. Hence, the force-free fields, obeying the G-S equation, on two sides of the light surface should be uncorrelated. From this perspective, it is reasonable to treat the light surface as a boundary, on the two sides of which the force-free fields should be solved independently. In other words, fields on two sides of the light surface could never interact unless the force-free condition is voilated. For instance, the appearrance of current sheet may correlate the force-free field across the light surface \cite{Camilloni:2022kmx}.

\section{Summary and Discussions}
\label{sec:sum}

In this work, we provide a general expression of the G-S equation, shown in Eq.~\eqref{eq:G-S-2}, which describes a force-free electromagnetic field with symmetries along an arbitrary pair of commuting vector fields in an arbitrary 4D manifold, via a second order PDE of the scalar potential $\psi$. A simplified expression is also derived in Eq.~\eqref{eq:G-S-4}, which is equivalent to Eq.~(86) in Ref~\cite{Gralla:2014yja} and is applicable as long as the 4D manifold is Riemannian and 2+2 decomposible. The correctness of Eq.~\eqref{eq:G-S-2} is verified by deriving from it the specific forms of the Grad-Shafranov equation in several well-known cases. The expression for the case of the stationary spacetime in 3+1 decomposition is also obtained. Furthermore, we argue that Eq.~\eqref{eq:G-S-2} could be viewed as the on-shell condition of scalar field $\psi$ whose dynamics is captured by the Lagrangian density given in Eq.~\eqref{eq:lagrangian}. Based on Eq.~\eqref{eq:lagrangian}, $\psi$ behaves as massless particles moving on hypersurfaces perpendicular to $\mathcal{Z}$, for which the light surface constitutes a natural boundary that these particles could only approach asymptotically.

As stated in Sect.~\ref{sec:equation}, ${\rm d}f\equiv e_3\cdot F  \neq 0$ is the prerequisite for the validity of Eq.~\eqref{eq:G-S-2}. The situation of ${\rm d}g=0$ corresponds to $\omega\equiv 0$, as illustrated by the first two examples in Sect.~\ref{sec:example}. On the other hand, the situation with ${\rm d}g\neq 0$ but ${\rm d}f=0$ could be handled by simply exchanging the roles of $e_0$ and $e_3$. For instance, in the first example discussed in Sect.~\ref{sec:example}, people may consider $\partial_{z}\cdot F=0$ instead of $\partial_t\cdot F=0$. We can then appropriately choose $e_0=\partial_{z}$ and $e_3=\partial_t$, with which all the calculations become valid. The resulting equation is identical to Eq.~\eqref{eq:G-S-carte} except that $\partial_x\psi=E_x$ and $\partial_y\psi=E_y$. Namely, $\psi$ now becomes the electric potential. When ${\rm d}f={\rm d}g=0$, meaning that both $e_0$ and $e_3$ annihilate ${\rm d}\phi_1$ and ${\rm d}\phi_2$, the field sheets lie solely in the subspace spanned by $e_0$ and $e_3$. Consequently, the field takes the form of $F=H{\rm d}\xi\wedge{\rm d}\zeta$, with the magnitude $H$ satisfying $d(H^2)=0$, indicating the energy conservation.

The force-free electrodynamics, in which the G-S equation plays a vital role, is effective for analyzing electromagnetically dominated environments such as the solar corona, tokamak, Earth's magnetosphere and black hole magnetosphere. We hope this work will help researchers quickly obtain the specific form of the G-S equation in a variety of complicated situations. For example, what if the symmetries of the field are not aligned with the Killing vectors of the spacetime? More specifically, what if the field is not stationary in the conventionally chosen rest frame? In recent years, inclined and precessing relativistic jets launched from rotating black holes have attracted increasing attention in astrophysics \cite{Liska:2017alm,Liska:2019svb,Chatterjee:2020eqc,Selvi:2024lsh}. However, most related studies rely on numerical methods. It may be possible to address this scenario using force-free electrodynamics, since the equation of equilibirum could be quickly derived from Eq.~\eqref{eq:G-S-2} once the coordinates $\{\tau,\chi,\xi,\zeta\}$ are suitably determined. Additionally, the coronal mass ejection in magnetized accretion flow near black hole has been widely studied \cite{Yuan:2008zv,Xi:2024sxb} and is considerably more complex than that in the solar system, with spacetime curvature introducing additional difficulties.. With Eq.~\eqref{eq:G-S-2}, deriving the equilibrium equation for coronal mass ejection near a black hole should at least become a tractable task.

\begin{acknowledgments}
I appreciate the constructive suggestions from the anonymous referee. I am grateful for the support of my parents. I would also like to thank Yuedan Wang for her constant encouragement.
\end{acknowledgments}

\appendix

\section{Differential Forms}
\label{sec:form}

\subsection{Definition and Basic Operations}
\label{sec:def}

A $p$-form is defined to be an antisymmetric covariant rank-$p$ tensor. The rank cannot be exceed the dimension of the manifold (or it is identically zero). A scalar is a 0-form while a covariant vector is a 1-form. The electromagnetic tensor is represented by a 2-form. Basic operations on differential forms includes wedge product, exterior derivative and inner product.

The wedge product “$\wedge$” turns a $p$-form and a $q$-form into a $(p+q)$-form as:
\begin{equation}
    \left(A\wedge B\right)_{\mu_1...\mu_p\nu_1...\nu_q}
    :=\frac{(p+q)!}{p!q!}A_{[\mu_1...\mu_p}B_{\nu_1...\nu_q]}
    \label{eq:wedge_def}
\end{equation}
where the square bracket is the operation of antisymmetrization. For any 1-form $A$, we obviously have $A\wedge A=0$. Exchanging the order of differential forms in wedge product satisfies:
\begin{equation}
    A\wedge B = (-1)^{pq}B\wedge A
    \label{eq:wedge_prop}
\end{equation}

Exterior derivative “${\rm d}$” produces a $(p+1)$-form after acting on a $p$-form like:
\begin{equation}
    ({\rm d}A)_{\rho\mu_1...\mu_p}:=(p+1)\nabla_{[\rho}A_{\mu_1...\mu_p]}
    \label{eq:d}
\end{equation}
The covariant derivatives (compatible to the metric) here could be replaced by any torsion-free derivative. Acting exterior derivative twice on any $p$-form yields zero:
\begin{equation}
    {\rm d}({\rm d}A)=0
    \label{eq:dd}
\end{equation}
which is apparent from its definition. The exterior derivative acting on a wedge product of a $p$-form and a $q$-form satisfies:
\begin{equation}
    {\rm d}(A\wedge B)={\rm d}A\wedge B + (-1)^p A\wedge{\rm d}B
    \label{eq:d_prop}
\end{equation}
which means the exterior derivative is graded. A $p$-form $F$ is called “closed” if ${\rm d}F=0$. It is called “exact” whenever there is a $(p-1)$-form $A$ satisfying ${\rm d}A=F$. It is provable that “closed” and “exact” are equivalent. The elctromagnetic tensor is a typical example of a closed and exact 2-form.

A $p$-form can be contracted with up to $p$ vectors, either from the left or right:
\begin{equation}
    \begin{aligned}
        &(v\cdot A)_{\mu_2...\mu_p} := v^{\mu_1}A_{\mu_1...\mu_p} \\
        &(A\cdot v)_{\mu_2...\mu_p} := v^{\mu_p}A_{\mu_1...\mu_p}
    \end{aligned}
    \label{eq:inner1}
\end{equation}
which defines the inner product between a $p$-form and a vector. One can also define the inner product between a rank-$q$ contravariant tensor and a $p$-form when $q\leq p$ as:
\begin{equation}
    (B\cdot A)_{\mu_{q+1}...\mu_p} := B^{\mu_1...\mu_q}A_{\mu_1...\mu_q...\mu_p}
    \label{eq:inner2}
\end{equation}
The square of a $p$-form is just the inner product of the $p$-form and its corresponding rank-$p$ contravariant tensor, which is not necessarily non-negative.

\subsection{Hodge Dual}
\label{sec:hodge}

In an $n$-dimensional manifold, the metric-compatible volume element is an $n$-form defined as:
\begin{equation}
    \epsilon_{\mu_1...\mu_p} := \sqrt{\left|g\right|}\varepsilon_{\mu_1...\mu_p}
    \label{eq:volume1}
\end{equation}
where $g$ is the determinant of metric tensor while $\varepsilon_{\mu_1...\mu_p}$ is the $n$-dimensional Levi-Civita symbols. It is feasible to define the metric-compatible volume element via the coordinate basis $\{e^a\}$ as:
\begin{equation}
    \epsilon := \sqrt{\left|g\right|}e^1\wedge...\wedge e^n
    \label{eq:volume2}
\end{equation}
Any $n$-form on $n$-dimensional manifold is proportional to the metric-compatible volume element. The normalization of metric-compatible volume element is given by:
\begin{equation}
    \epsilon^2=\epsilon\cdot\epsilon = (-1)^s n!
    \label{eq:e-dot-e}
\end{equation}
Here $s$ is the number of “$-1$” in the metric tensor expressed under orthonormal bases, referred to be the negative index of inertia of the matrix $[g_{\mu\nu}]$. For example, $s=0$ for Euclidean manifold and $s=1$ for Lorenzian manifold. 

The Hodge dual of a $p$-form in an $n$-dimensional manifold is a $(n-p)$-form define via the metric-compatible volume element as:
\begin{equation}
    \ast A := \frac{1}{p!}A\cdot\epsilon
    \label{eq:hodge}
\end{equation}
We call “$\ast$” the Hodge star operator. Acting Hodge star operator twice on a $p$-form results in itself up to a sign:
\begin{equation}
    \ast\ast A = (-1)^{s+p(n-p)}A
    \label{eq:hodge-square}
\end{equation} 
where $s$ is just that in Eq.~\eqref{eq:e-dot-e}. For any pair of $p$-forms, there is a useful formula:
\begin{equation}
    A\wedge\ast B = \frac{1}{p!}(A\cdot B) \epsilon
    \label{eq:hodge_prop}
\end{equation}

\subsection{Spacetime Decomposition}
\label{sec:decompose}

In physics, there are many cases that the metric tensors are block-diagonal, such as the Minkowski spacetime in Cartesian, cylindrical or spherical coordinates and the Kerr spacetime in Boyer-Lindquist coordinates. In these cases, it is convenient to decompose the manifold into several submanifolds. 

Let us go back to our main problem in this paper. We are discussing the electromagnetic field in a 4D spacetime with symmetries along two commuting 1-forms ${\rm d}\tau$ and ${\rm d}\chi$. It is often encountered that the metric tensor is block-diagonal, meaning that the 4D metric tensor can be expressed as a direct sum of two 2D metric tensor like:
\begin{equation}
    [g_{\mu\nu}]=[g^L_{\mu_1\nu_1}]\oplus[g^R_{\mu_2\nu_2}]=
    \begin{pmatrix}
        [g^L_{\mu_1\nu_1}] & {\bf 0} \\
        {\bf 0} & [g^R_{\mu_2\nu_2}]
    \end{pmatrix}
    \label{eq:decomposition}
\end{equation}
where $[g^L]$ denotes the metric tensor of the 2D submanifold (usualy Lorenzian) spanned by ${\rm d}\tau$ and ${\rm d}\chi$, while $[g^R]$ denotes the metric tensor of the 2D submanifold (usually Riemannian) spanned by ${\rm d}\xi$ and ${\rm d}\zeta$. The full 4D manifold is decomposed into two 2D submanifolds. As a consequence, the metric-compatible volume element could be decomposed as $\epsilon=\epsilon^L\wedge\epsilon^R$ with:
\begin{equation}
    \epsilon^L=\sqrt{\left|g^L\right|}{\rm d}\tau\wedge{\rm d}\chi~~~,~~~
    \epsilon^R=\sqrt{\left|g^R\right|}{\rm d}\xi\wedge{\rm d}\zeta
    \label{eq:volume3}
\end{equation}
where $g^L$ and $g^R$ denote the determinants of metric tensors respectively. The volume elements $\epsilon^L$ and $\epsilon^R$ are Hodge duals of each other up to a sign, which is determined by the properties of the coordinate basis.

\section{Degenerate Field}
\label{sec:degenerate}

A degenerate field is a field satisfying $F\wedge F=0$, or $F_{[\mu\nu}F_{\kappa\rho]}=0$ in elements. Classically, the condition is equivalent to $\Vec{E}\cdot\Vec{B}=0$, which means either the electric and magnetic strength are perpendicular to each other or any one of them vanishes in some certain rest frame.

In a 4D manifold, for an undetermined electromagnetic field $F$, assume that one can find a nonzero 4-vector $v$ perpendicular to the field, meaning that $F\cdot v=0$. Then the equation $\left(F\wedge F\right)\cdot v=0$ is subsequently satisfied. As we know, any 4-form in a 4D manifold is proportional to the 4D Levi-Civita tensor. So the only way to keep $\left(F\wedge F\right)\cdot v=0$ on any point of the manifold is to require $F\wedge F=0$. To sum up, an electromagnetic field in a 4D manifold is degenerate whenever there is a nonzero 4-vector orthogonal to the field everywhere. Force-free conditions indicates that the current is orthogonal to the field. Therefore, the force-free electromagnetic field is a typical example of degenerate field.

A degenerate field could be expressed by a wedge product of two 1-forms as:
\begin{equation}
    F=\alpha\wedge\beta
    \label{eq:AwB}
\end{equation}
It is easy to obtain $F\wedge F=0$ from $F=\alpha\wedge\beta$. To derive $F=\alpha\wedge\beta$ from $F\wedge F=0$, let us consider two nonzero unparalleled 4-vectors $v$ and $w$ satisfying $\Tilde{\alpha}=F\cdot v$ and $\Tilde{\beta}=F\cdot w$. When computing $0=v\cdot\left(F\wedge F\right)\cdot w$, one can only dig out two kinds of terms, which are $\Tilde{\alpha}\wedge\Tilde{\beta}$ and $\left(w\cdot v\cdot F\right)F$ (omit the signs). As long as the scalar $\left(w\cdot v\cdot F\right)$ is nonzero, one can choose, for example, $\alpha=\left(w\cdot v\cdot F\right)^{-1}\Tilde{\alpha}$ and $\beta=\Tilde{\beta}$ so that $F=\alpha\wedge\beta$ is satisfied. 

A degenerate electromagnetic field has two degree of freedom \cite{Uchida1,Uchida2}. Since the field is closed, it is resultantly satisfied that
\begin{equation}
    {\rm d}\alpha\wedge\alpha\wedge\beta ={\rm d}\beta\wedge\alpha\wedge\beta=0
    \label{eq:frobenius} 
\end{equation}
The set of equation above is just the Frobenius condition which confirms the complete integrability of $\alpha$ and $\beta$ \cite{Chernbook}. Namely, both $\alpha$ and $\beta$ are exact. As a consequence, a degenerate electromagnetic field could be decomposed by two scalars, called Euler potentials, as $F={\rm d}\phi_1\wedge{\rm d}\phi_2$ \cite{Uchida1,Uchida2}. It is a generalization of the notion that the magnetic (or electric) field could be expressed by a bunch of 1D unintersecting smooth lines, which could be viewed as intersections of two piles of smooth 2D slices. The intersections of constant $\phi_1$ and $\phi_2$ are called the field sheets. The choice of $\{\alpha,\beta\}$, hence $\{\phi_1,\phi_2\}$, is not unique. One can definitely make the substitution $\{\phi_1,\phi_2\}\rightarrow\{\Tilde{\phi}_1,\Tilde{\phi}_2\}$ without changing the field as long as:
\begin{equation}
    \left|\frac{\partial\left(\Tilde{\phi}_1,\Tilde{\phi}_2\right)}{\partial\left(\phi_1,\phi_2\right)}\right|=1
\end{equation}
Namely, the unit Jacobian of the substititions confirms that the two different sets of Euler potentials describe one degenerate field.

With Eq.~\eqref{eq:AwB}, the force-free condition becomes:
\begin{equation}
    \left(\alpha\wedge\beta\right)\cdot j=\left(\beta\cdot j\right)\alpha-\left(\alpha\cdot j\right)\beta=0
    \label{eq:ff1}
\end{equation}
where $j$ is the current 4-vector (or 1-form). There are two apparent choices for the satisfaction of Eq.~\eqref{eq:ff1}. First, $\alpha$ and $\beta$ are proportional to each other. In this case, however, the electromagnetic field is identically zero, which points to a trivial situation. Second, $\alpha\cdot j=0$ and $\beta\cdot j=0$ are simultaneously satisfied. These two equations are equivalent to $\alpha\wedge J=0$ and $\beta\wedge J=0$, where $J:=\epsilon\cdot j$ is the current 3-form. Consequently, the force-free condition could be expressed by two equations of 4-forms in 4D manifold (hence two equations of scalars equivalently) as shown in Eq.~\eqref{eq:ff}.

\section{Comoving 1-form}
\label{sec:comoving}

Let us restrict our discussion in a 4D Lorenzian spacetime, with timelike $e_0=\partial_{\tau}$ and spacelike $e_3=\partial_{\chi}$. One can define an observer moving along $e_3$ as $u\propto e_0+v_3e_3$, for $v_3$ being the magnitude of 3-velocity (or angular velocity) of the observer. We know that any spacelike vector annihilating $u$ could be taken as a static ruler in the view of this observer. When $v_3=\omega$, for which $u\equiv L(e_0,e_3)$, it is satisfied that $u\cdot\mathcal{Z}\equiv 0$. Therefore, the 1-form $\mathcal{Z}$ could be regarded as a ruler moving along $e_3$ with the magnitude of 3-velocity to be $\omega$. 

Recall that $u$ annihilates both ${\rm d}\phi_1$ and ${\rm d}\phi_2$ when $v_3=\omega$ so that $u\cdot F=0$. It means the observer is stuck at the magnetic field line (another statement of frozen theorem \cite{Gralla:2014yja}) and there is no electric field in the view of $u$. As a consequence, the motion of $u$ along $e_3$ is simultaneous to the transverse shift of magnetic field line. In this sense, one can regard $\omega$ to be the velocity of the field. The timelike property of $u$ is kept if and only if $\mathcal{Z}$ is spacelike. While on the critical surfaces where $\mathcal{Z}$ becomes null, $u$ becomes null as well (hence it can never be interpreted as an observer). It is reasonable to explain that a critical surface is a 2D subspace on which the field moves in the speed of light. That is why we usually call the critical surfaces to be the light surfaces.

\section{Existence of Unique Solution}
\label{sec:unique}

A linear second order partial differential equation (PDE) of function $\psi\left(x_i\right)$ takes a general form as:
\begin{equation}
    \sum_{i,j=1}^n \mathcal{A}^{ij}\frac{\partial^2\psi}{\partial x_i \partial x_j}
    +\sum_{i=1}^n \mathcal{B}^i \frac{\partial\psi}{\partial x_i}+\mathcal{C}\psi =0
    \label{eq:gen-linear}
\end{equation}
where $\mathcal{A}^{ij}$, $\mathcal{B}^{i}$ and $\mathcal{C}$ are real smooth functions of $\{x_i\}$ on an $n$-dimensional region $U$ with boundary $\partial U$. The coefficients of second order partial derivative terms make up a characteristic matrix $[\mathcal{A}^{ij}]$ of the PDE, called the Hessian matrix. Eq.~\eqref{eq:gen-linear} is referred to be elliptic if its Hessian matrix of is positive-definite for any point on $U$. Based on the Sylvester criterion, Eq.~\eqref{eq:gen-linear} is elliptic if and only if:
\begin{equation}
    \mathcal{A}^{11}>0~,~
    \left|
    \begin{matrix}
        \mathcal{A}^{11} & \mathcal{A}^{12} \\
        \mathcal{A}^{21} & \mathcal{A}^{22}
    \end{matrix}
    \right|>0~,~\ldots~
    \left|
    \begin{matrix}
        \mathcal{A}^{11} & \cdots & \mathcal{A}^{1n} \\
        \vdots & \ddots & \vdots \\
        \mathcal{A}^{n1} & \cdots & \mathcal{A}^{nn}
    \end{matrix}
    \right|>0
    \label{eq:sc}
\end{equation}
Namely, the sequential principal minors of Hessian matrix are required to be positive for an elliptic PDE. 

One useful theorem for a linear elliptic PDE could be stated as followed. When $\mathcal{C}\leq 0$, after providing the Dirichlet boundary condition on $\partial U$, Eq.~\eqref{eq:gen-linear} has a unique solution on $U$ if it is elliptic, satisfying (i) the solution $\psi\left(x_i\right)$ is differentiable to the second order on $U$ and (ii) $\psi\left(x_i\right)$ is continuous on $U+\partial U$ \cite{courant89,Evans}. 

The discussion in the linear PDE could be extended to the non-linear case. A non-linear second order PDE could be expressed generally as:
\begin{equation}
    K\left(x_i,\psi,\psi_{i},\psi_{ij}\right)=0
    \label{eq:gen-nonlinear}
\end{equation}
where $K$ is a real differentiable function on an n-dimensional region $U$ with boundary $\partial U$. Here we denote $\partial\psi/\partial x_i$ as $\psi_{i}$ and $\partial^2\psi/\left(\partial x_i \partial x_j\right)$ as $\psi_{ij}$ for abbreviation. The Hessian matrix of Eq.~\eqref{eq:gen-nonlinear} is defined as:
\begin{equation}
    \left[\mathcal{A}^{ij}\right]:=\left[\frac{\partial K}{\partial \psi_{ij}}\right]
    \label{eq:hessian}
\end{equation}
and Eq.~\eqref{eq:gen-nonlinear} as called elliptic as long as its Hessian matrix is positive-definite. The theorem of unique solution for a non-linear second order PDE is stated as follows. The second order PDE expressed by Eq.~\eqref{eq:gen-nonlinear} has a unique solution on $U$ after provding the Dirichlet boundary condition on $\partial U$ as long as (i) it is elliptic and (ii) $\partial K/\partial \psi \leq 0$. Additionally, the unique solution $\psi\left(x_i\right)$ is differentiable to the second order on $U$ and is continuous on $U+\partial U$. 

If the function $\psi$ depends on two variables $\{x_1,x_2\}$ only, the Hessian matrix of Eq.~\eqref{eq:gen-nonlinear} is a $2\times 2$ square matrix. In this case, the conditions for the existence of unique solution are equivalent to:
\begin{equation}
    \frac{\partial K}{\partial \psi_{11}}\frac{\partial K}{\partial \psi_{22}}-\left(\frac{\partial K}{\partial \psi_{12}}\right)^2>0~~,~~
    \frac{\partial K}{\partial \psi_{11}}\frac{\partial K}{\partial \psi}\leq 0
    \label{eq:unique}
\end{equation}

It is noteworthy that the conditions introduced above are sufficient but not necessary for the existence of unique solution. In other words, the violation of the conditions does not mean the second order PDE has no solution or its solution is not unique.

Now, focusing on the second order derivative terms, let us reorganize Eq.~\eqref{eq:G-S-2} to be:
\begin{equation}
    \left(\left|\mathcal{Z}\right|^2g^{\beta\gamma}-\mathcal{Z}^{\beta}\mathcal{Z}^{\gamma}\right)\partial_{\beta}\partial_{\gamma}\psi+\ldots=0
    \label{eq:elliptic-GS}
\end{equation}
Henceforth, the Hessian matrix of the G-S equation should be:
\begin{equation}
    \left[\mathcal{A}^{\beta\gamma}\right]\equiv \left[\left|\mathcal{Z}\right|^2g^{\beta\gamma}-\mathcal{Z}^{\beta}\mathcal{Z}^{\gamma}\right]
\end{equation}
which is a $2\times 2$ square matrix since $\{\beta,\gamma\}$ could be $\{\xi,\zeta\}$ only. Meanwhile, 
the expression of $\mathcal{Z}$ indicates:
\begin{equation}
    \frac{\partial \mathcal{Z}_{\alpha}}{\partial \psi}=-\frac{d\omega}{d\psi}\left({\rm d}\tau\right)_{\alpha}~~,~~
    \frac{\partial\left(\partial_{\alpha}\mathcal{Z}_{\beta}\right)}{\partial \psi}=-\partial_{\alpha}\psi\frac{d^2\omega}{d\psi^2}\left({\rm d}\tau\right)_{\beta}
    \label{eq:Z-psi}
\end{equation}
Only $G$ and $\omega$ are explicitly depend on $\psi$, so the factor $\partial K/\partial \psi$ in the second equation of Eq.~\eqref{eq:unique} should be:
\begin{equation}
    \begin{aligned}
        \frac{\partial K}{\partial \psi}=&\frac{(-1)^s}{4}\left|{\rm d}\tau\wedge{\rm d}\chi\right|^2\frac{d^2\left(G^2\right)}{d\psi^2} \\
        &+Y_{21}\frac{d^2\omega}{d\psi^2}-2Y_{11}\frac{d\omega}{d\psi}+Y_{12}\left(\frac{d\omega}{d\psi}\right)^2
    \end{aligned}
    \label{eq:K_psi}
\end{equation}
The coefficients on the second line read:
\begin{equation}
    \begin{aligned}
        Y_{21}=&\left(\mathcal{Z}\cdot\partial\psi\right)\left({\rm d}\tau\cdot\partial\psi\right)-\left(\mathcal{Z}\cdot{\rm d}\tau\right)\left|\partial\psi\right|^2 \\
        Y_{11}=&\left(\mathcal{Z}\cdot{\rm d}\tau\right)\left[\left(\partial\ln\sqrt{\left|g\right|}\cdot\partial\psi\right)+\partial\cdot\partial\psi\right] \\
        &-\mathcal{Z}^{(\beta}\left({\rm d}\tau\right)^{\gamma)}\left[\partial_{\beta}\left(\ln\sqrt{\left|g\right|}\right)\partial_{\gamma}\psi+\partial_{\beta}\partial_{\gamma}\psi\right]\\
        Y_{12}=&\left|{\rm d}\tau\right|^2\left|\partial\psi\right|^2-\left({\rm d}\tau\cdot\partial\psi\right)^2
    \end{aligned}
\end{equation}

Consider that one would like to solve the G-S equation shown in Eq.~\eqref{eq:G-S-2} on the region $U$ with boundary $\partial U$. Based on the theorem discussed in Appen.~\ref{sec:unique}, after providing the Dirichlet boundary condition on $\partial U$, the G-S equation exists a unique solution on $U$ whenever the following two conditions are satisfied:
\begin{equation}
    \begin{aligned}
        {\rm (i):}~~&{\rm Det}\left(\left[\big|\mathcal{Z}\big|^2g^{\beta\gamma}-\mathcal{Z}^{\beta}\mathcal{Z}^{\gamma}\right]\right)>0 \\
        {\rm (ii):}~~&\frac{\partial K}{\partial \psi}\left[\left|\mathcal{Z}\right|^2g^{\xi\xi}-\left(\mathcal{Z}^{\xi}\right)^2\right]\leq 0
    \end{aligned}
    \label{eq:ss1}
\end{equation}
where ${\rm Det}\left(...\right)$ denotes the determinant of a square matrix. In other words, one could judge the existence and uniqueness of the solution according to the conditions exhibited in Eq.~\eqref{eq:ss1} before solving the G-S equation. One set of Dirichlet boundary condition on $\partial U$ determines at least one and only one global structure of the field described by Eq.~\eqref{eq:G-S-2} in $U$ as long as Eq.~\eqref{eq:ss1} is satisfied.

The conditions shown in Eq.~\eqref{eq:ss1} are actually not user-friendly. However, simplications could be made in some cases of interest. For example, if we consider the field in a decomposable 4D manifold as introduced in Sect.~\ref{sec:simplification}, the first condition in Eq.~\eqref{eq:ss1} reduces to:
\begin{equation}
    \big|\mathcal{Z}\big|^4\left(g^R\right)^{-1}>0
    \label{eq:ss21}
\end{equation}
Mostly, we have $\left(g^R\right)^{-1}>0$. The first condition is satisfied as long as $\mathcal{Z}$ is not a null 1-form. That is, we are not considering the field on the light surface. If we restrict $\omega\equiv 0$ on every point of $U$ furtherly, the second condition in Eq.~\eqref{eq:ss1} becomes:
\begin{equation}
    (-1)^{s+s_l}g^{\xi\xi}\left|g^L\right|^{-1}\left|\mathcal{Z}\right|^2\frac{d^2\left(G^2\right)}{d\psi^2}\leq 0
    \label{eq:ss22}
\end{equation}
where $s_l$ is the negative index of inertia of $[g^L_{\mu\nu}]$. We mostly have $s=s_l=1$ and $g^{\xi\xi}>0$. Henceforth, for a spacelike $\mathcal{Z}$, the existence of unique solution needs a convex $G^2(\psi)$. While for a timelike $\mathcal{Z}$, a concave $G^2(\psi)$ is required.

\section{Proofs}
\label{sec:tags}

\subsection{Proof of Eq.~\eqref{eq:dfdg}}
\label{sec:prove-eq4}

Here we prove the second step of Eq.~\eqref{eq:dfdg}. For two vectors ${X,Y}$ and one 2-forms $F$, based on the definition of dot products, we have:
\begin{equation}
    X\cdot F \equiv X^{\mu}F_{\mu\nu}~~~,~~~Y\cdot F \equiv Y^{\mu}F_{\mu\nu}= -Y^{\nu}F_{\mu\nu}
\end{equation}
The right-hand-sides give the expressions by elements. Hence their wedge product yields: 
\begin{equation}
    \left(X\cdot F\right)\wedge\left(Y\cdot F\right) \equiv 
    -2X^{\mu}F_{\mu[\nu}F_{\rho ]\kappa}Y^{\kappa}
\end{equation}
We know that $F$ is internally antisymmetric. With the operation of antisymmetrization, we have: $F_{\mu[\nu}F_{\rho ]\kappa} = -F_{\mu[\nu}F_{\kappa ]\rho} = F_{\mu[\kappa}F_{\nu ]\rho}$. It thus indicates:
\begin{equation}
    \begin{aligned}
        \left(X\cdot F\right)\wedge\left(Y\cdot F\right) &\equiv 
        -2X^{\mu}F_{\mu[\nu}F_{\rho ]\kappa}Y^{\kappa} \\
        &= 2X^{\mu}F_{\mu[\kappa}F_{\nu ]\rho}Y^{\kappa} \\ 
        &\equiv\left(Y\cdot X\cdot F\right)F
    \end{aligned}
\end{equation}
which is just the second step of Eq.~\ref{eq:dfdg}.

\subsection{Terms of ${\rm d}\phi_2\wedge J$}
\label{sec:term-vanish}

We already have:
        ${\rm d}\phi_2 = {\rm d}\eta+\mathcal{Z}-\omega'\tau{\rm d}\psi$ and
        $J = G'{\rm d}\psi\wedge{\rm d}\tau\wedge{\rm d}\chi+{\rm d}\ast\left({\rm d}\psi\wedge\mathcal{Z}\right)$.
The vanishing of $\omega'\tau{\rm d}\psi\wedge\left(G'{\rm d}\psi\wedge{\rm d}\tau\wedge{\rm d}\chi\right)$ and $\mathcal{Z}\wedge\left(G'{\rm d}\psi\wedge{\rm d}\tau\wedge{\rm d}\chi\right)$ are apparent. We are going to prove $\omega'\tau{\rm d}\psi\wedge{\rm d}\ast\left({\rm d}\psi\wedge\mathcal{Z}\right)={\rm d}\eta\wedge{\rm d}\ast\left({\rm d}\psi\wedge\mathcal{Z}\right)=0$. Since $\psi$ is independent of $\{\tau,\chi\}$, we may conveniently set: ${\rm d}\psi = P{\rm d}\xi + Q{\rm d}\zeta$, where $P\equiv\partial_{\xi}\psi$ and $Q\equiv\partial_{\zeta}\psi$. $P$ and $Q$ depend only on $\{\xi,\zeta\}$ as well. The the second term of $J$ could be expressed as:
\begin{widetext}
\begin{equation}
    \begin{aligned}
        {\rm d}\ast\left({\rm d}\psi\wedge\mathcal{Z}\right) &\propto 
        {\rm d}\ast\left( \pm P{\rm d}\xi\wedge{\rm d}\chi \pm P\omega{\rm d}\xi\wedge{\rm d}\tau \pm Q{\rm d}\zeta\wedge{\rm d}\chi \pm Q\omega{\rm d}\zeta\wedge{\rm d}\tau \right) \\
        &\propto {\rm d}\left( \pm P{\rm d}\zeta\wedge{\rm d}\tau \pm P\omega{\rm d}\zeta\wedge{\rm d}\chi \pm Q{\rm d}\xi\wedge{\rm d}\tau \pm Q\omega{\rm d}\xi\wedge{\rm d}\chi \right) \\
        &\propto {\rm d}\xi\wedge{\rm d}\zeta\wedge\left( \pm \frac{\partial P}{\partial\xi}{\rm d}\tau \pm \frac{\partial\left(P\omega\right)}{\partial\xi}{\rm d}\chi  \pm \frac{\partial Q}{\partial\zeta}{\rm d}\tau \pm \frac{\partial\left(Q\omega\right)}{\partial\zeta}{\rm d}\chi \right)
    \end{aligned}
    \label{eq:expr}
\end{equation}
\end{widetext}
where “$\pm$” indicates “$+$” or “$-$” as the signs are irrelavant to the consequence. From Eq.~\eqref{eq:expr}, we know that the second term of $J$ contains the factor of ${\rm d}\xi\wedge{\rm d}\zeta$. Additionally, since both the 1-forms $\omega'\tau{\rm d}\psi$ and ${\rm d}\eta$ are linear combinations of ${\rm d}\xi$ and ${\rm d}\zeta$, it is henceforth inevitable that $\omega'\tau{\rm d}\psi\wedge{\rm d}\ast\left({\rm d}\psi\wedge\mathcal{Z}\right)={\rm d}\eta\wedge{\rm d}\ast\left({\rm d}\psi\wedge\mathcal{Z}\right)=0$

One can imitate the calculations shown above to examine that the second and third terms on the right-hand-side of Eq.~\eqref{eq:ff1-case1} are identically zero.

\nocite{*}

\bibliography{refs}

\end{document}